\documentclass[
  reprint,
 superscriptaddress,
 amsmath,amssymb,
 aps,
]{revtex4-2}

\usepackage{graphicx}
\usepackage{dcolumn}
\usepackage{bm}
\usepackage{latexsym}
\usepackage{xcolor}
\usepackage{svg}
\usepackage{natbib}
\usepackage{cmbright}
\usepackage[font=sf,justification=justified]{caption}
\usepackage[font=small,labelfont=bf,
   justification=Justified,
   format=plain]{caption}

\usepackage{color}

\begin{document}
\title{Locomotion of a Scallop-Inspired Swimmer in Granular Matter}

\author{Hongyi Xiao}
\affiliation{Institute of Multiscale Simulation, Friedrich-Alexander-Universität Erlangen-Nürnberg, Erlangen, Germany}
\affiliation{Department of Mechanical Engineering, University of Michigan, Ann Arbor, USA}%
\author{Harol Torres}
\affiliation{Institute of Multiscale Simulation, Friedrich-Alexander-Universität Erlangen-Nürnberg, Erlangen, Germany}
\affiliation{Sioux Technologies GmbH, Erlangen, Germany}
\author{Achim Sack}
\affiliation{Institute of Multiscale Simulation, Friedrich-Alexander-Universität Erlangen-Nürnberg, Erlangen, Germany}
\author{Thorsten Pöschel}
\affiliation{Institute of Multiscale Simulation, Friedrich-Alexander-Universität Erlangen-Nürnberg, Erlangen, Germany}

\date{\today}

\begin{abstract}
Understanding swimming in soft yielding media is challenging due to their complex deformation response to the swimmer's motion. We experimentally show that a scallop-inspired swimmer with reciprocally flapping wings generates locomotion in granular matter. This disagrees with the scallop theorem prohibiting reciprocal swimming in a liquid when its inertia is negligible. We use X-ray tomography and laser profilometry to show that the propulsion is created by the combined effects of jamming and convection of particles near the wings, which break the symmetry in packing density, surface deformation, and kinematics of the granular medium between an opening and a closing stroke.
 \end{abstract}

\maketitle

\noindent Understanding locomotion in complex media is key to a variety of applications involving animals and robotics in soft and yielding terrains \cite{hosoi2015beneath}, microorganisms and micromachines in biofluids \cite{lauga2011life, li2017micro, hu2018soft, gao2018targeting, arratia2022life}, etc. Despite significant progress \cite{lauga2011life, hosoi2015beneath, arratia2022life}, key questions remain unanswered regarding how the rheology of such media interacts with a swimmer's body deformation. Even with the simplest swimming motions, such as reciprocal motions involving identical body deformation sequences when time is reversed, complex physics can emerge in complex media. \cite{normand2008flapping, keim2012fluid, qiu2014swimming, texier2021propulsion}. 

Swimming at low Reynolds numbers with reciprocal motion is generally impossible due to negligible fluid inertia, as argued by Purcell in the ``scallop theorem'' \cite{purcell1977life}. In the case of complex fluids, especially polymer fluids, the rheological properties of the medium can lead to propulsion \cite{arratia2022life, lauga2011life}. A scallop-like swimmer that opens and closes at different rates can propel itself in shear thickening or thinning liquid \cite{qiu2014swimming}, and swimmers generating curved streamlines can utilize the elasticity of the medium to generate locomotion \cite{keim2012fluid, normand2008flapping}.
While the fundamental principles of fluid mechanics apply to locomotion in viscoelastic liquids \cite{normand2008flapping} with the swimmer size being orders of magnitude larger than the fluid's molecules, locomotion in granular matter can be harder to understand \cite{li2013terradynamics, hosoi2015beneath, aguilar2016review} due to the lack of scale separation in granular matter \cite{IGscales, TanGoldhirsch:1998, Goldhirsch:2008} and the dissipative nature or particle interaction leading to liquid-like and solid-like behaviors \cite{jaeger1996granular, andreotti2013granular} depending on the excitation.

Granular materials can experience a jamming transition under increasing confining pressure or shear stress \cite{liu1998jamming, bi2011jamming}, and a jammed granular matter behaves like a disordered solid \cite{nicolas2018deformation, cubuk2017structure, Holger24}. Under small strain, it responds elastically, while under large strain, it deforms plastically with particle rearrangements where the yield is often heterogeneous with strain concentrating in localized shear bands \cite{schall2010shear}. The jamming transition has important implications for granular locomotion. Particles near an intruding object tend to jam and, thus, effectively enlarge the object \cite{brzinski2013depth, aguilar2016robophysical, harrington2020stagnant, kang2018archimedes}. The mechanical response of two nearby intruders is non-additive, suggesting long-range interaction \cite{pacheco2010cooperative,pravin2021effect, agarwal2021efficacy, carvalho2022collaborative}. 
The response of granular matter often depends on the history of previous deformations. Therefore, an intruder such as an undulating snake may experience different resistance when encountering the same region twice \cite{schiebel2020mitigating}.

In granular matter, propulsion can be achieved through non-reciprocal and complicated mechanisms \cite{hosoi2015beneath}, such as undulation for slender swimmers \cite{maladen2009undulatory, hatton2013geometric, goldman2014colloquium, schiebel2019mechanical}, fluidization \cite{shimada2009swimming,naclerio2021controlling}, and adding compliance to flapping swimmers \cite{li2021compliant,chopra2023toward}. Here, we ask whether simple and reciprocal motion, considered by the \textit{scallop theorem}, can cause locomotion in granular matter. A recent experimental study showed that a rod-shaped intruder that oscillates around one of its ends, driven by an external device, can propel itself in a granular material \cite{texier2021propulsion}. A phenomenological model was proposed based on the hysteresis of the torque and the material relaxation, but the underlying physics was not fully understood.

\begin{figure}[t]
\includegraphics[width=1.0\linewidth]{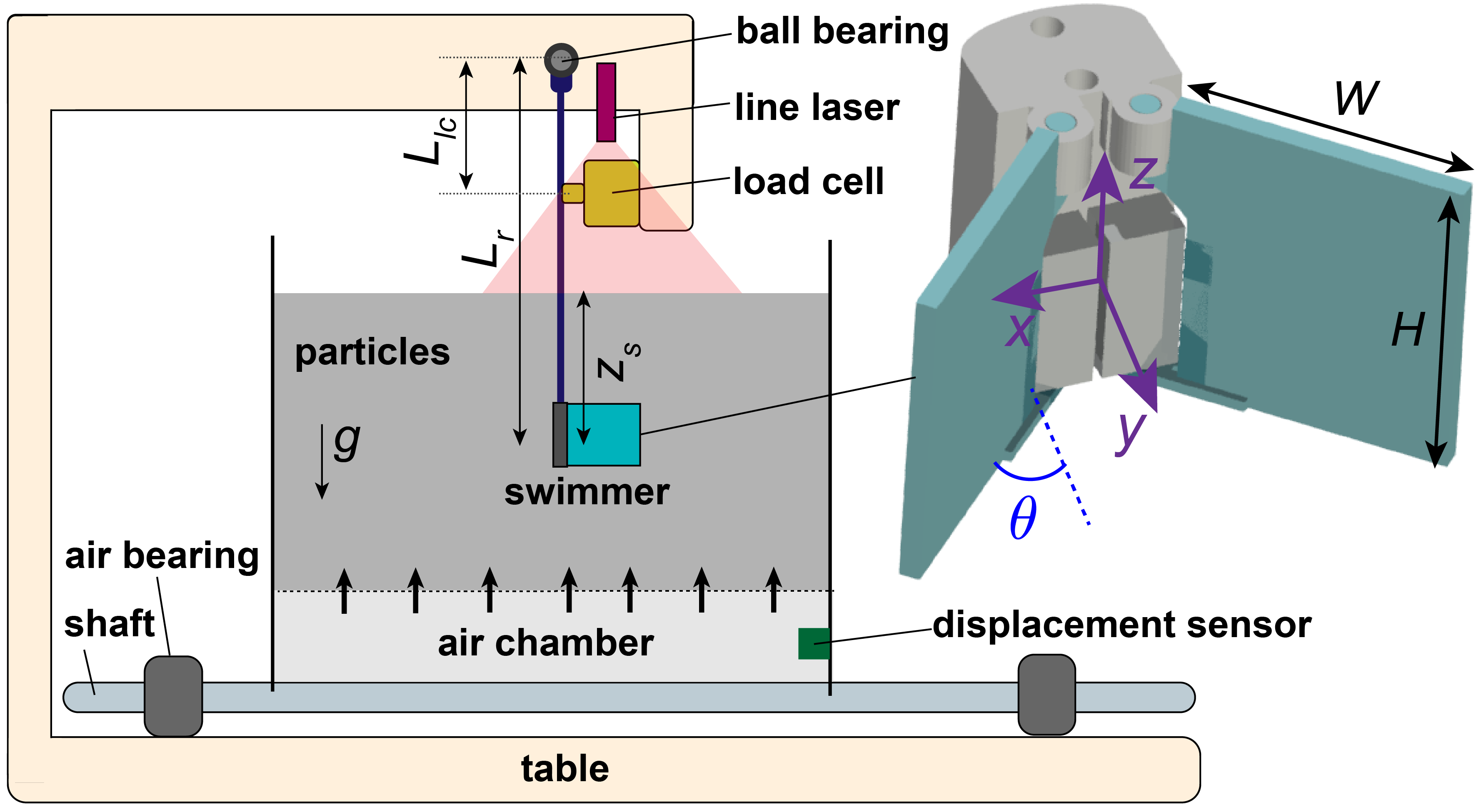}
\caption{\label{fig:exp} Experimental setup to study reciprocal swimming in a basin filled with granular material. A load cell records the force acting on the swimmer and a displacement sensor records the translation of the reservoir. LASER profilometry reveals the shape of the free surface. 
}
\vspace{-5pt}
\end{figure}

We present experiments using a scallop-inspired swimmer with two wings that open and close reciprocally, moving in a basin filled with granular material (Fig. \ref{fig:exp}). In contrast to the oscillating rod \cite{texier2021propulsion}, the swimmer does not need external support since its counter-moving wings, driven by two servo motors, balance the torque arising from the interaction of the wings with the granular material. Thus, the swimmer can move autonomously in the granulate without external support, similar to animals crawling in the soil. The rotational motion of the plate-shaped wings induces strong jamming effects that can change the rheological behavior of the surrounding medium. The oscillation of the swimmer's wings can cause deformation of the free surface, which is also observed for near-surface swimmers \cite{trouilloud2008soft, schiebel2020mitigating}. Based on three-dimensional (3d) experiments using LASER profilometry and X-ray computed tomography (CT), we reveal the mechanisms leading to the propulsion and discuss two symmetry-breaking mechanisms due to the formation of a stagnation zone in the granulate and due to granular convection. 

\vspace{+5pt}
\noindent \textbf{Scallop-like swimmer generates locomotion}

\noindent In the experimental setup shown in Fig. \ref{fig:exp}, the swimmer has two rectangular-shaped wings with width $W=12d$, height $H=11d$, mounted at a distance of $3.4d$ to each other. The thickness of the wings is $0.9d$. The ambient granulate consists of graphite-coated polydisperse expanded polystyrene particles with a diameter of $d=(4.1\pm 0.5)\,\text{mm}$ and a density of $\rho=22\,\text{kg/m}^3$. A rod fixes the swimmer's position, which is $25d$ away from both the initial free surface and the bottom of the reservoir. The swimmer's coordinates $(xyz)$ are relative to the reservoir, where $z$ is the vertical direction, $y$ is the swimming direction, and $x$ is the lateral direction, perpendicular to $z$ and $y$. By construction, $z$ and $x$ are invariant during the experiment.

The granular reservoir is placed on a sliding rail with air bearings to eliminate translational friction. Therefore, the swimmer's motion can be determined from the reservoir's position. The force felt by the swimmer in the $y$ direction, $F_y$, is measured by a load cell connected to the rod. In each experiment, we first air-fluidize the reservoir to refresh the particle packing structure and create a flat free surface \cite{jinpreparation,qian2015dynamics}, see Sec. Methods. Then, 
the wings open and close periodically in reciprocal motion at constant angular velocity $\omega=30^\circ\text{s}^{-1}$ in the interval $\theta_\text{o}\le \theta\le \theta_\text{c}$ with the period $T=2(\theta_o-\theta_c)/\omega$.

\begin{figure}[t]
\includegraphics[width=1.0\linewidth]{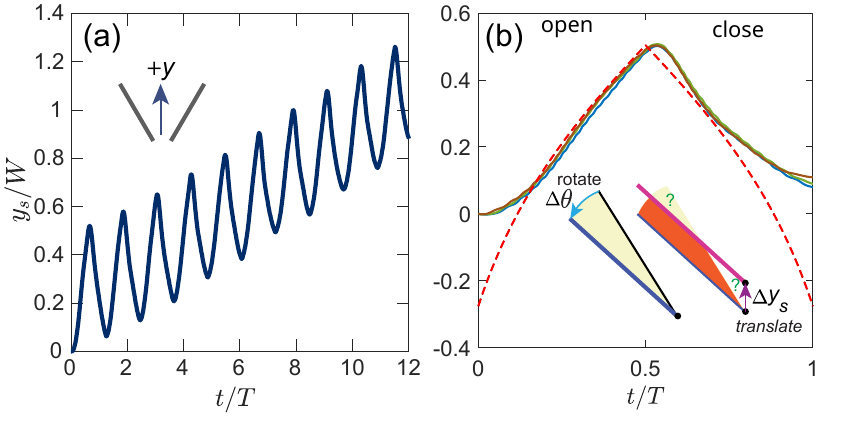}
\caption{\label{fig:free} Reciprocal free swimming in granular matter with the wings oscillating between $\theta_c=20^\circ$ and $\theta_o=80^\circ$. (a) Swimmer displacement as a function of time. The inset indicates the direction of the motion. 
(b) Same data as (a), but the displacement is drawn over one cycle for each period individually. The dashed line is the integral of Eq. \eqref{eq:2}, shifted to match the peak of the measurement. The inset illustrates the areas swept by a step wing rotation and translation. 
}
\vspace{-15 pt}
\end{figure}

Figure \ref{fig:free}a shows the displacement of the swimmer relative to the reservoir, $y_s$, as a function of time. Figure \ref{fig:free}b shows the same data, but the displacement is drawn for each period individually (for clearness, only 3 periods are shown). We observe a uniform motion of the swimmer in the positive $y$-direction, with minor deviations between individual periods. We also noticed that the swimming direction hardly depends on the swimmer's depth and frequency $\omega$ in the accessible range; the latter is consistent with the known fact that granular rheology is rate-independent at low deformation rate \cite{maladen2009undulatory,hatton2013geometric}.

A simple geometrical argument is considered here. Neglecting inertia, two volumes are relevant: a small wing rotation, $\Delta \theta$, will displace particles of volume $V_R=\frac{1}{2}\Delta\theta W^2H$, and the induced swimmer translation, $\Delta y_s$, will displace particles of volume $V_T=\Delta y_s W\sin{\theta}H$, at given wing angle $\theta$. These volumes are illustrated in the inset of Fig \ref{fig:free}b. If $V_T$ fully compensates $V_R$, we obtain

\vspace{-15 pt}
\begin{equation}
\frac{\Delta (y_s/W)}{\Delta\theta}= \frac{1} {2 \sin{\theta}}\,,
\label{eq:2}
\end{equation}

\noindent which can be integrated over a swimming cycle to obtain the swimmer's displacement. Representing $\theta$ as a function of time, Eq. \eqref{eq:2} predicts the scaled swimmer's displacement as a function of time, $y_s/W=f(t/T)$, shown by the dashed line in Fig. \ref{fig:free}b.
The simple volume balance, Eq. \eqref{eq:2}, predicts the swimmer's translation well at large angles ($0.2<t/T<0.7$), but it deviates considerably from the measured displacement for small angles, $t/T\lesssim 0.2$ and particularly when closing, $t/T\gtrsim 0.7$. 

Two deformation mechanisms of the surrounding granular medium can affect the validity of Eq. \eqref{eq:2}: the change of density of the granular material near the swimmer, induced by the swimmer's motion and the lateral and vertical convection of particles around the swimmer's wings. Both effects are, in turn, influenced by granular jamming that makes particles near the swimmer hard to rearrange. These effects are difficult to observe as the swimmer is submerged, and slow swimming generates only a small net force. We, therefore, modified the described experiment in two different ways to enable precise force measurement and 3d imaging.

\vspace{+5pt}
\noindent \textbf{Probing jamming effects via clamped swimming}

\begin{figure}[t]
\includegraphics[width=1.0\linewidth]{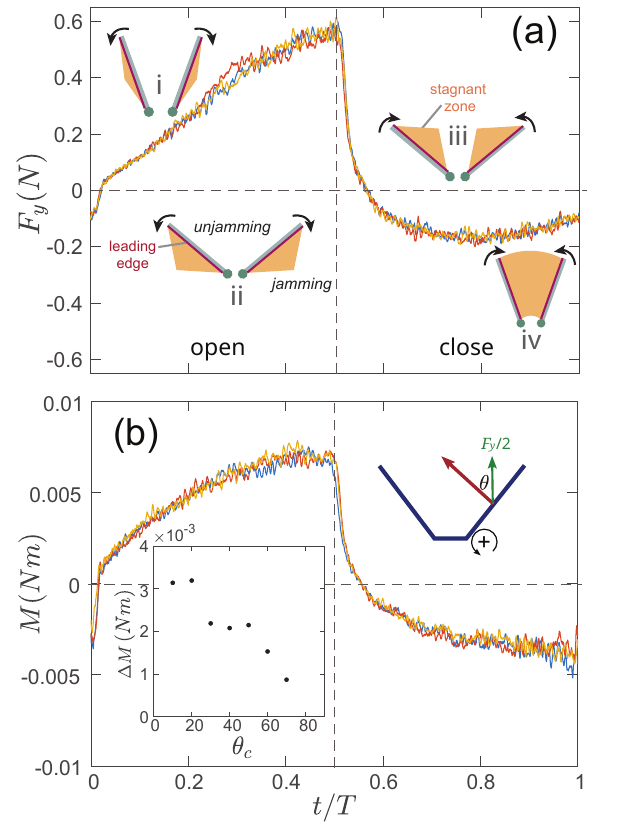}
\caption{\label{fig:jamming} Asymmetry of the propulsion force in clamped swimming. (a) Propulsion force vs. time for $\theta_c=20^\circ$ and $\theta_c=80^\circ$.
The inset sketches the stagnant zone formation before the leading edges at different time instants.
(b) The corresponding torque exerted by a swimmer. Three individual cycles for either case are shown. 
The inset shows the difference in peak values between open and close vs. the close angle.
}
\vspace{-20pt}
\end{figure}

\noindent To identify jamming effects from the propulsion force, we modified the experiment: we immobilized the reservoir by blocking the air bearings and, thus, fixed the swimmer's position relative to the granulate. This is a common approach to measure locomotion (e.g. \cite{normand2008flapping, qiu2014swimming, saro2023hydrodynamic}). We measure the resulting propulsive force during the swimmer's oscillations, $F_y$. In the following, we call the experiment with variable relative distance, in which we measure the displacement as a function of time, ``free swimming''. The experiment with constant distance, in which we measure the force as a function of time (or angle $\theta$), will be referred to as ``clamped swimming''. Figure \ref{fig:jamming}a shows the propulsive force vs. time for three individual periods. We find that the forces of the opening and closing strokes are significantly asymmetric. A net positive cycle-averaged propulsion force in the positive $y$-direction is evident, in agreement with the free-swimming experiment (Fig. \ref{fig:free}). 

For a pure rotational motion, it is more appropriate to consider the resistive torque, $M$, felt by the motor than $F_y$, which is shown in Fig. \ref{fig:jamming}b as a function of the normalized time $t/T$. 
The relation between the torque and the force reads $M=\frac{1}{4}WF_y/\sin\theta$, with the sign convention sketched in Fig. \ref{fig:jamming}b for the right wing. When the wings approach the largest opening angle ($t/T\approx0.5$), the resistance of the granular medium against the wing rotation is also highest, resulting in a large force, $F_y$, in the swimming direction. Near the maximum closing angle ($t/T\approx1$), the granulate's resistance against rotation is also large. Nevertheless, the resulting force, $F_y$, is small as the wings mostly push towards the $x$ direction. This geometrical difference results in a positive cycle-averaged propulsion force. 

The increase in $M$ as a function of $\theta$ suggests the formation of stagnant zones where particles are densely packed and are thus unable to rearrange and flow relative to each other. Stagnant zones near flat intruders (such as the swimmer's wings) and their relation to granular jamming have been reported, e.g., in \cite{muller2014granular, harrington2020stagnant}.
The insets of Fig. \ref{fig:jamming}a sketch the formation of stagnant zones for different phases of the swimmer's oscillation. The wing's leading edge that pushes against the stagnant zone is highlighted in red. Stagnant zones consisting of jammed granulate have the properties of a disordered solid and, thus, effectively enlarge the wings. Their occurrence and shape differ for the opening and closing half-cyle as the leading edge switches sides. This asymmetry eventually causes asymmetry of the forces and, thus, locomotion.
Another indicator of jamming can be seen in the delay between the direction change of $M$ and the direction change of the wing's rotation at $t/T=0$ and $t/T=0.5$. Such a delay would not occur for a wing rotating slowly in an ambient viscous liquid.

In Fig. \ref{fig:jamming}b, the peak magnitude of $M(t)$ at the maximum opening is bigger than that at the maximum closing. We further examine this difference in a series of clamped experiments with $\theta_\text{o}=80^\circ$, and $\theta_\text{c}$ ranging from 10$^\circ$ to 70$^\circ$. The difference between the peak values, $\Delta M=\left|M(\theta_\text{o})\right| - \left|M(\theta_\text{c})\right|$, is shown in the inset of Fig. \ref{fig:jamming}b. The difference $\Delta M$ is always positive and a decreasing function of $\theta_c$. This behavior is due to vertical convection, as described below. 

\vspace{+5pt}
\noindent \textbf{Surface deformation due to vertical convection}

\begin{figure}[t]
\includegraphics[width=1.0\linewidth]{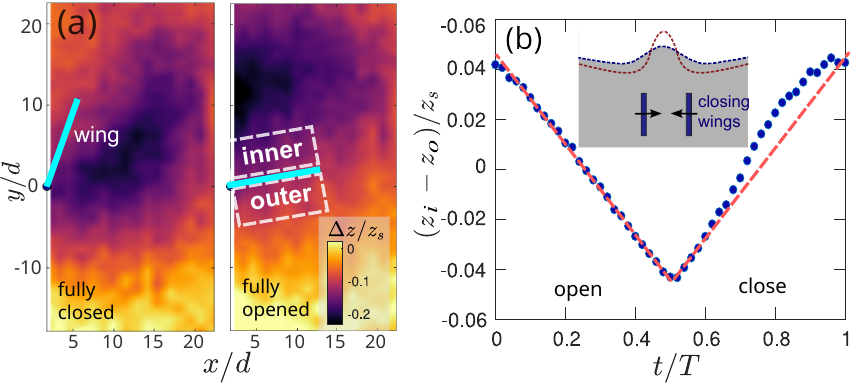}
\caption{\label{fig:surface} Surface deformation measured with LASER profilometry for $\theta_\text{c}=20^\circ$ and $\theta_\text{o}=80^\circ$. (a) Difference in height relative to the initial surface height, $\Delta z_s$, in the $xy$-plane in the vicinity of the swimmer, for $\theta=20^\circ$ (left) and $\theta=80^\circ$ (right). The cyan line shows the position of the submerged wing. (b) difference of surface height averaged in regions in front of and behind the wing versus $t/T$. White boxes in panel (a) indicate the area used for averaging. The dashed lines are symmetric about $t/T=0.5$. The inset sketches the surface deformation during the closing stroke. 
}
\end{figure}

\noindent 
The flow of particles in front of the leading edges is hindered by jamming, and the only direction the wing can drive the particles is upwards, causing vertical convection and free surface deformation.
Earlier studies have shown that the heaps and valleys formed by a submerged intruder's motion influence the resistive force \cite{gravish2010force, schiebel2020mitigating, jin2020small, agarwal2021surprising}. For our system, the formation of surface structures due to the swimmer's motion is shown in Fig. \ref{fig:surface}a. The color codes the height of the free surface, $\Delta z$, with respect to the initial surface height, $z_s=25d$, for a clamped swimmer (only the right half of the domain is shown). An overall negative $\Delta z$ can be seen due to the swimmer pumping particles to the $-y$ direction as well as possible densification from the as-fluidized initial condition. Closing the wings pushes material toward the front of the swimmer, creating a heap in front of the swimmer and a valley behind it when the wings are fully closed. During the opposite stroke, the surface flow is reversed, which results in a reverse height difference when the wings are fully open. 

For a more quantitative analysis, we averaged $\Delta z$ over the inner (front) region and outer (back) region with respect to the wings. The white boxes in 
Fig. \ref{fig:surface}a indicate the regions used for averaging. Figure \ref{fig:surface}b shows the height difference relative to the homogeneous initial height, $(z_\text{i}-z_\text{o})/z_\text{s}$, as a function of time, $t/T$, over a cycle. The development of the height difference with each stroke is consistent with the increase of $|M|$ during each stroke, as shown in Fig. \ref{fig:jamming}b.

The measured relative height difference, $(z_\text{i}-z_\text{o})/z_\text{s}$ is not symmetric with respect to $t/T=1/2$. Significant asymmetry occurs for $t/T\gtrsim 0.7$ when the two wings are in close proximity to each other. Consequently, the deformation of the granulate caused by the swimmer's motion is not reciprocal between opening and closing. The process is sketched in Fig. \ref{fig:surface}b: when the wings are in close proximity near the end closing stroke, a heap forms (red dotted line) since displaced material accumulates in the narrow region between the wings. When the slope of the heap exceeds the angle of repose, particles on the surface flow downhill, shallowing the heap (blue dotted line). The opening process creates two less steep and distant heaps, one for each wing. This difference in heap building between opening and closing can contribute to the positive $\Delta M$ in Fig. \ref{fig:jamming}b. 

\vspace{+5pt}
\noindent \textbf{Density change and granular convection}

\begin{figure}[t]
\includegraphics[width=0.95\linewidth]{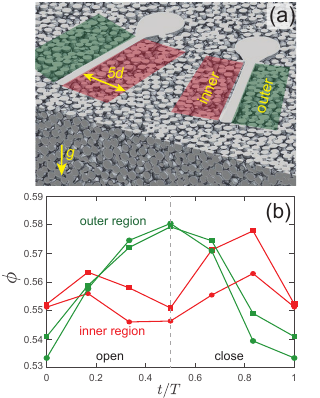}
\caption{\label{fig:density} X-ray CT results for a free swimmer with quasi-static wing rotation between $\theta_c=15^\circ$ and $\theta_o=70^\circ$. (a) an example reconstructed 3d image. (b) packing density at different cycle time instants averaged over inner regions (red in (a)) and outer regions (green in (a)). Circle and square symbols show the results from two repetitions. The data points at $t/T=1.0$ repeat the measurement at $t/T=0$.  
}
\end{figure}

\begin{figure*}
\includegraphics[width=1.0\linewidth]{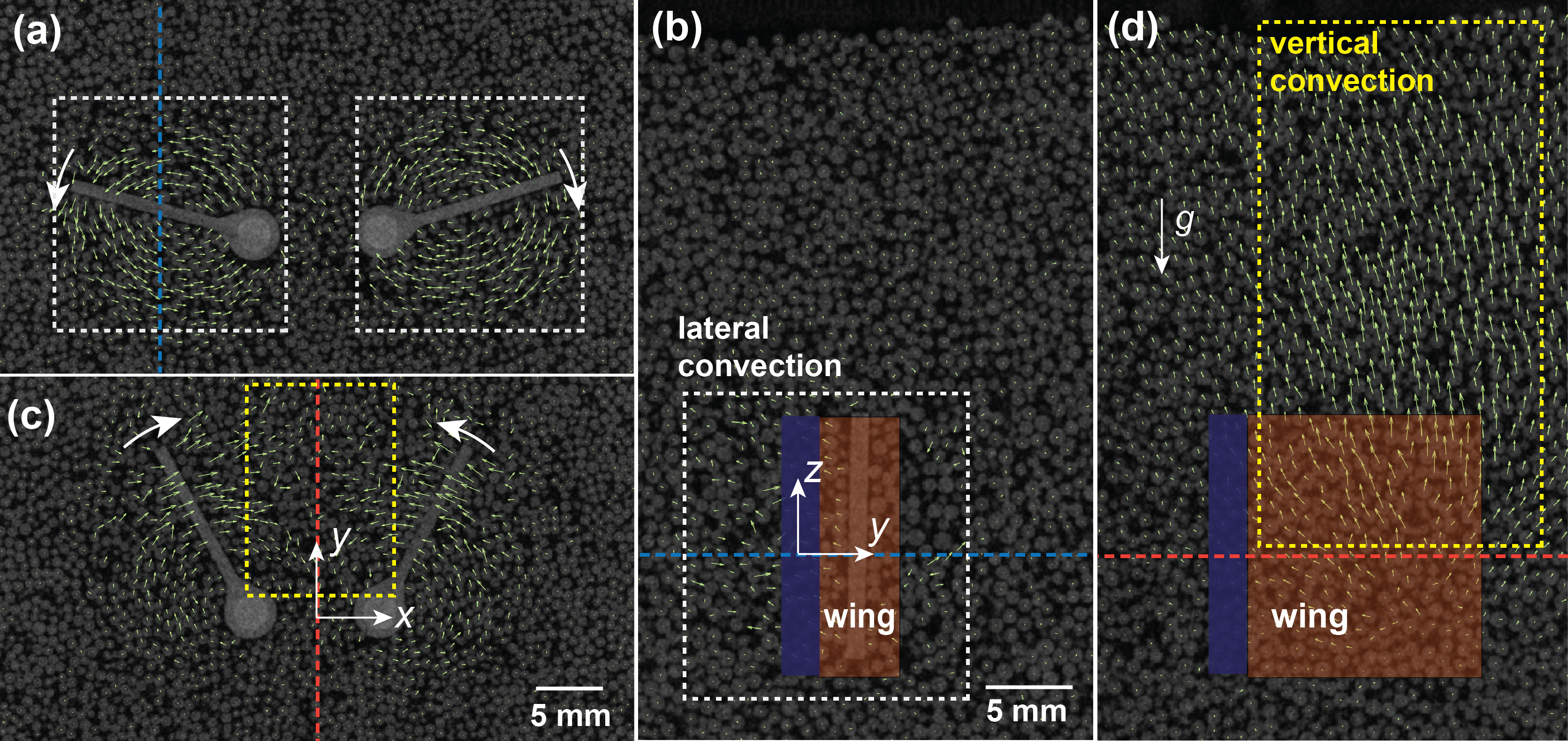}
\caption{\label{fig:xray} Cut through X-ray tomograms of the miniature free swimmer in the $xy$ and $yz$ planes with $\theta_c=20^\circ$ and $\theta_o=80^\circ$. (a,b) show tomograms for $\theta=75^\circ$ during opening, corresponding to $t/T=0.46$. (c,d) show tomograms for $\theta=25^\circ$ during closing, corresponding to $t/T=0.96$. (b) shows the $yz$ plane at the midpoint of the left wing, indicated by the blue dashed line in (a). (d) shows the $yz$ plane at the center of the swimmer, indicated by the red dashed line in (c). The orange and violet regions indicate the projections of the wing and the anchoring rod on the $yz$ plane. The white and yellow boxes highlight regions with strong horizontal and vertical convection, respectively. The yellow arrows indicate the displacements (magnified by 2$\times$) of the particles between consecutive frames, recorded with delay $T/30$.}
\end{figure*}

\noindent With the jamming mechanism observed in clamped swimming in mind, we now consider the more complicated free swimming problem. To this end, we performed the experiment at a slightly smaller scale inside an X-ray CT, which allows for imaging of the 3d structure of the granulate. From this information, we can determine the packing density and the field of flow velocity in the vicinity of the swimmer. The setup of the experiment is identical to the one described (free swimming) except for the smaller size of the wings ($\text{1.5}\times \text{1.5\, cm}^2$) of the swimmer that moves with reciprocal motion in a reservoir filled with 1\,mm polystyrene beads. The ratio between particle sizes and wing size was kept similar. 

We performed the experiment quasi-statically and stopped the wings' motion in steps of $6^o$ to record a series of high-resolution tomograms covering the entire cycle, see Sec. \textit{Methods} for more details. From these 3d tomograms, we determine the position of the grains and, subsequently, the field of packing density (volume fraction), $\phi$. Figure \ref{fig:density}a shows a horizontal cut through such a tomogram. We average the packing density in volumes ($W\times H\times 5d$) in the inner and the outer regions of the wings, as indicated in Fig. \ref{fig:density}a. Figure \ref{fig:density}b shows two independent measurements of the averaged densities as a function of $t/T$. 

We produced 30 tomograms in equidistant steps of $T/30$ to measure the field of particle velocities, see a video in the supplementary. The velocities are then obtained from the displacements of the particles. Figure \ref{fig:xray}a and c show cuts along the $xy$ plane. The center of the swimmer gives the $z$ coordinate. The arrows show the projection of the particle displacement vectors between two consecutive frames onto the $xy$ plane. Following the same method, we show particle displacement projected on to $yz$ planes in Fig. \ref{fig:xray}b and d. The $x$ position of the planes are indicated by the dashed lines in Fig. \ref{fig:xray}a and b, which were chosen to highlight vertical convection while the majority of particles remain in-plane. Particles without an associated arrow experience out-of-plane motion.


The opening stroke, Fig. \ref{fig:xray}a, is characterized by the formation of a stagnant zone in front of the leading edge, which is the outer region in Fig. \ref{fig:density}. Here the material is jammed, that is, the material is of high density such that the particles can hardly move relative to each other but follow the wing's rotation coherently, which can be seen in Fig. \ref{fig:xray}a. The development of this stagnant zone is characterized by an increase in torque, see Fig. \ref{fig:jamming}a, and an increase in density in the outer region in Fig. \ref{fig:density}b during opening. The granular convection at large opening angles occurs laterally in the $xy$ plane, as indicated in the white boxes in Fig. \ref{fig:xray}a, where particles in front of the leading edge tend to flow coherently from the tip to the anchor of the wings. Hardly any vertical particle motion can be seen in the region immediately above the swimmer Fig. \ref{fig:xray}b. This means that the volume balance for wing rotation and translation required for Eq. \eqref{eq:2} can be easily met, resulting in the good agreement between the slope in Fig. \ref{fig:free}b. This also means that the granular convection is highly localized and the swimmer only need to displace a small volume of particles to move forward.


A similar stagnant zone can be found during the closing stroke, shown in Fig. \ref{fig:xray}c. 
The inner region between the wings is now pushed by the new leading edge of the wing and its density increases during closing in Fig \ref{fig:density}. Near the end of closing, a strong upward convection occurs reaching all the way to the free surface, which is observed in the highlighted region in the $yz$ plane in Fig. \ref{fig:xray}d. Correspondingly, in Fig. \ref{fig:xray}c, we lost track of many particles in that region due to out-of-plane motion, while particles immediately next to the wings still have coherent in-plane displacement due to jamming effect. The occurrence of strong vertical convection coincides with the period when a significant lack of backward displacement is seen in Fig. \ref{fig:free}b, a drop of density in the inner region near the end of closing in Fig. \ref{fig:density}b, and a lack of resistive torque in Fig. \ref{fig:jamming}c. 

In this case, the balance of volumes leading to Eq. \eqref{eq:2} needs to be extended to $V_R=V_T+V_E$, in which $V_E$ is the volume of vertically escaped particles in a step rotation. This will result in the smaller swimmer translation at small angles when $V_E$ is large. As the stagnant zones are compressible, this balance also needs to include the how the packing density $\phi$ changes as shown in \ref{fig:density}b. As $\phi$ is asymmetric in the inner region between opening and closing, the symmetry between swimmer translation during opening and closing is broken. 



\vspace{+5pt}
\noindent \textbf{Conclusions}

\noindent 
The scallop-inspired swimmer can generate propulsion in an ambient granular material. Even though its movement is entirely reciprocal, the response of the granulate is not. Employing LASER profilometry and x-ray tomography, we showed that the asymmetric response of the granulate results from a combination of granular flow due to jamming and granular flow due to surface relaxation. The effects studied here are rate-independent, provided the simmer's movement is slow compared to the time scale of surface relaxation, as the jamming transition occurs at a much shorter time scale. For more rapid movement of the swimmer, more complicated effects can be expected for three reasons: (a) the time scale of the surface relaxation interferes with the time scale of the swimmer's movement, (b) the rate-dependent granular rheology \cite{gonzalez2009reciprocal,hubert2021scallop}, and (c) the swimmer's inertia. The study of these effects is beyond the experimental techniques used here.


\renewcommand\thefigure{\thesection A\arabic{figure}}  
\setcounter{figure}{0}  

\vspace{+5pt}
\noindent \textbf{Methods}

\vspace{+5pt}
\noindent \textbf{Experimental setup}

\noindent The experimental apparatus is sketched in Fig. \ref{fig:exp}: It consists of two rectangular wings driven by two servo motors located at a lateral distance of 13.8\,mm. The wings have a height of $H=52.0\,\text{mm}$, a width of $W=47.5\,\text{mm}$, and a thickness of 3.5\,mm. The swimmer is attached to a metal rod of length $L_\text{r}=24\,\text{cm}$, which allows the swimmer to be immersed in the particle reservoir. The upper end of the rod is connected to a ball bearing, and a load cell at a distance of $L_{\text{lc}}=5\,\text{cm}$ is used to measure the propulsion force. 

To create an initial packing with a reproducible packing density, the granular particles reservoir is fluidized with an upward air flow \cite{jinpreparation,qian2015dynamics} and then settle due to defluidization, achieving an initial particle volume fraction of $0.57\pm 0.003$. Further tapping of the reservoir can increase the density to $0.59\pm 0.004$. We have verified that this does not qualitatively change the swimmer's behavior \cite{texier2021propulsion}. 
The force felt by the swimmer in $y$-direction was obtained from the load cell signal, $F_{\text{lc}}$, via $F_y=F_{\text{lc}}L_\text{r}/L_{\text{lc}}$. A Hall-effect sensor measured the displacement of the reservoir, $y_{\text{re}}$, to an precision of $25\,\mu\text{m}$. Both signals have been measured at a rate of 10\,kHz, and a moving average with a window of 5\,ms was applied. LASER profilometry \cite{gravish2010force,takizawa2020novel} was used to monitor the height profile of the granulate's free surface with four line LASERs and a camera that captures images at 60\,Hz.
The angle of the wings, $\theta$, was measured by a rotary encoder built into the motor.  

\vspace{+5pt}
\noindent \textbf{Miniaturized experiment for X-ray imaging}

\noindent The miniaturized experiment was built according to the same design principles as the main experiment. Since the whole experiment was to be carried out inside a CT scanner, the mechanical construction was designed in such a way that only X-ray-transparent materials were used within the radiation area. With a pair of counter-rotating wings, the active swimmer moves submerged in a reservoir of solid polystyrene particles with a material density of $1050\,\text{kg/m}^{3}$. The motors were separated from the swimmer to ensure the radiographic transparency of the reservoir. The motors are located above the reservoir and drive the wings using very thin shafts. The geometry of this experiment maintains the ratio between the swimmer's wings ($15\times 15\,\text{mm}^2$), the mean particle diameter (1\,mm), the reservoir cross-section ($70\times 70\,\text{mm}^2$), and the swimmer's depth under the surface invariant. We applied the same air-fluidization protocol as described for the main experiment to create reproducible initial packings. We also used the same linear rail and air bearings for the free swimming experiment. We checked that the miniaturized experiment shows the same phenomena and characteristics as the main experiment.

\begin{figure}[h]
\includegraphics[width=\linewidth]{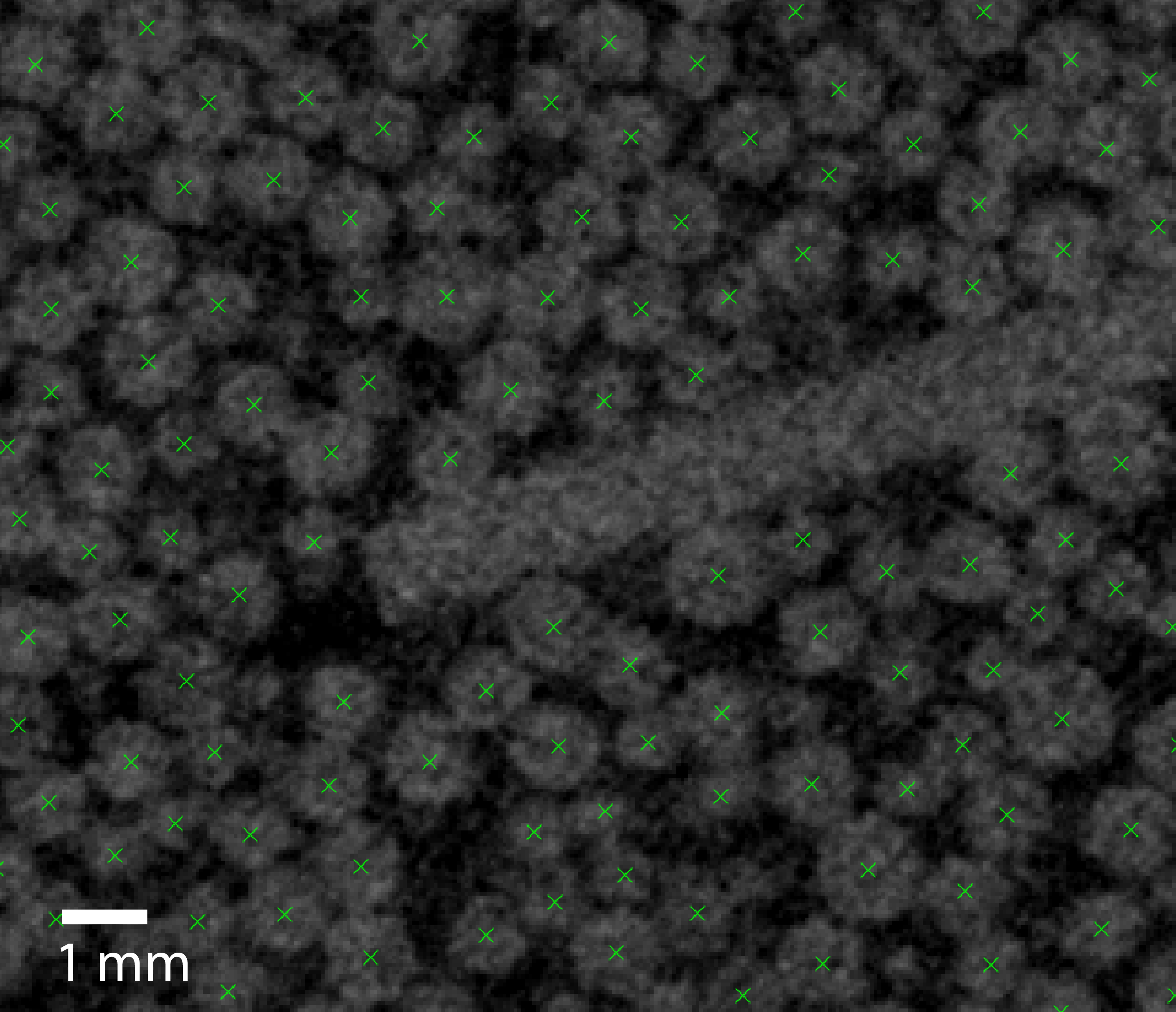}
\caption{Determining the particle positions with a neural network \cite{dillavou2024bellybutton}. The figure shows a horizontal cut along the $xy$ plane through a 3d tomogram. The green crosses indicate the determined particle locations.}
\label{fig:track}
\vspace{-15pt}
\end{figure}

We studied the swimmer's behavior at different velocities of the wing's rotation to obtain rate-independent results for $T>16\,\text{s}$, which determines the oscillation frequency for a quasi-static experiment. For the recording in the tomograph, we interrupted the experiment in steps of $6^\circ$. After each interruption, we recorded 1200 radiograms of detector resolution $1,936\times 1,536\, \text{pixel}^2$ to compute a full 3d tomogram. Figure \ref{fig:track} shows an example tomogram (horizontal slice of the 3d image). We used a neural network \cite{dillavou2024bellybutton} to determine the centers of the granular particles with the result demonstrated in Fig. \ref{fig:track}. 
We computed the field of particle displacement from the found particle locations for consecutive recordings (distance $6^\circ$ of the wing's motion). 

\vspace{+5pt}
\noindent \textbf{Acknowledgements}

\noindent We thank S. H. Ebrahimnazhad Rahbari, Jing Wang, Ralf Stannarius, Amir Nazemi, and Paul Umbanhowar for helpful discussions.

\noindent \textbf{Author contributions}

\noindent H.T. conceived the project. T.P. supervised the project and secured funding. H.T. and A.S. prototyped the experiment. H.X. performed the experiments, analyzed the results, and wrote the manuscript. All authors discussed the results and revised the manuscript.

\noindent \textbf{Competing interests}

\noindent The authors declare no competing interests.

\noindent \textbf{Data avalability}

\noindent The data reported in this study are available upon request to the corresponding author.


\def\bibsection{\section*{\refname}}
\bibliographystyle{unsrtnat}
\bibliography{swim.bib}

\end{document}